\documentclass[preprint]{aastex}
\shortauthors{CHUNG et al.}
\shorttitle{SEVERE DEGENERACIES IN BINARY LENSING EVENTS}

\newcommand{\uvec}{\mbox{\boldmath$u$}}
\newcommand{\thetavec}{\mbox{\boldmath$\theta$}}
\newcommand{\te}{t_{\rm E}}

\newcommand{\thetae}{\theta_{\rm E}}
\newcommand{\uthf}{u_{01,\rm {th}}}
\newcommand{\uths}{u_{02,\rm {th}}}
\newcommand{\alphao}{\alpha_{\rm o}}
\newcommand{\alphas}{\alpha_{\rm s}}

\begin{document}
\title{A New Application of the Astrometric Method to Break
Severe Degeneracies in Binary Microlensing Events}

\author{
Sun-Ju Chung\altaffilmark{1},
Byeong-Gon Park\altaffilmark{1},
Yoon-Hyun Ryu\altaffilmark{2}, and
Andrew Humphrey\altaffilmark{1}
}
\altaffiltext{1}{Korea Astronomy and Space Science Institute, Hwaam-Dong,
Yuseong-Gu, Daejeon 305-348, Korea; sjchung;bgpark;ajh@kasi.re.kr}
\altaffiltext{2}{Department of Astronomy and Atmospheric Science,
Kyungpook National University, Daegu 702-701, Korea;
yhryu@knu.ac.kr}


\begin{abstract}
When a source star is microlensed by one stellar component of a widely separated binary stellar components, after finishing the lensing event, the event induced by the other binary star can be additionally detected.
In this paper, we investigate whether the close/wide degeneracies in binary lensing events can be resolved by detecting the additional centroid shift of the source images induced by the secondary binary star in wide binary lensing events.
From this investigation, we find that if the source star passes close to the Einstein ring of the secondary companion, the degeneracy can be easily resolved by using future astrometric follow-up observations with high astrometric precision.
We determine the probability of detecting the additional centroid shift in binary lensing events with high magnification.
From this, we find that the degeneracy of binary lensing events with a separation of $\lesssim 20.0$ AU can be resolved with a significant efficiency.
We also estimate the waiting time for the detection of the additional centroid shift in wide binary lensing events. We find that for typical Galactic lensing events with a separation of $\lesssim 20.0$ AU, the additional centroid shift can be detected within $100$ days, and thus the degeneracy of those events can be sufficiently broken within a year.
\end{abstract}


\section{INTRODUCTION}

Observations are currently being undertaken for the detection of microlensing events, with survey observations being carried out by the OGLE \citep{udalski03} and MOA \citep{bond02} groups, and with follow-up observations being performed by the PLANET \citep{albrow01} and MicroFun \citep{dong06} groups.
The follow-up observations are focusing on high magnification events toward the Galactic bulge, so that they are efficient for detecting events with the characteristic features of binary lens systems, such as the caustic-crossing appearing in high magnification events. Moreover, if future space missions such as \emph{Microlensing Planet Finder} (\emph{MPF} \citep{bennett04}) with high photometric precision and monitoring cadence are carried out, the efficiency will significantly increase, and thus many more binary lensing events with the characteristic features will be found.

Detection of binary microlensing events is important for providing various information about the lensing parameters.
First, the binary mass ratio and projected separation are obtained. If many binary
lensing events are detected, these parameters can help to infer statistical properties of binaries in the lens population \citep{gaudi00}.
Second, the special case of caustic-crossing events provides the caustic-crossing time, $\Delta t_{cc}$, which is given by
\begin{equation}
\Delta t_{cc} = {\rho_\star \over |\sin\phi|} {\te} = {\theta_\star \over \mu|\sin\phi|}\ ,
\end{equation}
where $\rho_\star$ is the source radius normalized to the Einstein ring radius, corresponding to the total mass of the binary, $\te$ is the Einstein timescale of the binary lens, $\theta_\star$ is the angular source radius, $\phi$ is the caustic-crossing angle, and $\mu$ is the lens-source proper motion.
The angular source radius can be measured from its color and apparent magnitude.
The Einstein timescale and caustic-crossing angle can be determined from fitting of the observed light curve.
Third, we can obtain the information of the lens-source proper motion and the normalized source radius from equation (1).
Fourth, because $\mu = {\thetae / \te}$, if the proper motion is determined, the angular Einstein ring radius, $\thetae$, can be measured.
The Einstein ring radius is related to the total mass of the binary lens, $M$, which is expressed as
\begin{equation}
\thetae = \sqrt {{4GM \over c^2}{\left ({1 \over D_{\rm L}} - {1 \over D_{\rm S}}\right )}}\ ,
\end{equation}
where $D_{\rm L}$ and $D_{\rm S}$ are the distances to the lens and the source from the observer, respectively. Therefore, measuring the proper motion is very important to determine the lens mass and distance to the lens.

\citet{dominik99a} first reported that there exist multiple solutions to a photometrically observed binary lensing event.
\citet{han99} investigated the pattern of the centroid shift of the source images selecting four of the multiple solutions in the paper of \citet{dominik99a}.
As a result, all four cases showed different centroid shift patterns, and thus they demonstrated that the photometric degeneracy could be astrometrically resolved.
\citet{gould00}, however, showed that for the MACHO 98-SMC-1 event \citep{afonso00}, both of the light curves and centroid shift patterns of the close and wide binary solutions are extremely similar. The degeneracy of the close and wide binaries is rooted in the similarity of the lens equations of the two solutions, and thus it is very difficult to find a unique solution \citep{dominik99b}.
Nevertheless they found that the centroid shift curves of the two solutions have an offset at the point of the caustic-crossing time, and if the offset is observable through a space mission such as \emph{Space Interferometry Mission PlanetQuest} (\emph{SIM PlanetQuest} \citep{unwin08})
with high astrometric precision, a severe degeneracy could be astrometrically resolved.

In addition to the method of \citet{gould00}, there is a further method for breaking severe close/wide degeneracies of binary microlensing events.
For a wide binary lensing event, if a source star is lensed by one of the binary components (the first event), after this lensing event finishes, the event induced by the other binary star (the second event) can also be detected.
In this paper, we investigate whether the close/wide degeneracies in binary microlensing events can be resolved by detecting the centroid shift of the second event in wide binary lensing events.
We determine the probability of detecting the additional centroid shift in high magnification binary lensing events.
We also estimate the waiting time for the detection of the centroid shift of the second event after the caustic-crossing of the first event in wide binary lensing events.

The paper is organized as follows. In \S\ 2, we describe the astrometric properties of microlensing. In \S\ 3, we investigate a new astrometric method for breaking the degeneracy of binary lensing events and determine the probability of detecting the additional centroid shift in binary lensing events with high magnification.
In addition, we estimate the waiting time for the detection of the additional centroid shift in wide binary lensing events.
We conclude with a summary of our results and discussion in \S\ 4.

\section{ASTROMETRIC PROPERTIES OF MICROLENSING}

When a source star is lensed by a single lens, it is split into two images with individual magnifications and the centroid of the images is defined by
\begin{equation}
\theta_c = {1\over 2}{\left [ {u(u^2+4) \over {u^2 + 2}} + u \right ]}\thetae\ ;\ \ \
\uvec = {\left ( t-t_0 \over \te \right) {\hat{\it x}}+ {u}_{0}}\ {\hat{\it y}}\ ,
\end{equation}
where $u$ represents the projected lens-source separation in units of $\thetae$. The lensing parameters $u_0$ and $t_0$ are the lens-source impact parameter in units of $\thetae$ and the time of the maximum magnification, respectively.
The $x$ and $y$ axes represent the directions that are parallel with and normal to the lens-source transverse motion.
The centroid shift of the images with respect to the source star is determined by
\begin{equation}
\delta \thetavec_c = {\thetae \over {u^2 + 2}}\uvec\ .
\end{equation}

For a widely separated binary lensing event, each of the binary lens components works as the individual single lenses \citep{an02}. The centroid shift of this event can be described as the superposition of events induced by the individual lens components, and thus is expressed as
\begin{equation}
\delta \thetavec_c \sim \delta \thetavec_{c,1} + \delta \thetavec_{c,2}\ ,
\end{equation}
where $\delta \thetavec_{c,1}$ and $\delta \thetavec_{c,2}$ are the centroid shifts of the individual lens components, respectively.

The condition for wide binary lensing events to become photometric double lensing events is that the source should pass in the Einstein ring of the lens causing the second event. However, astrometric double lensing events can occur even though the source passes away from the Einstein ring of the lens. Therefore, the range of detecting the astrometric double lensing events is wider than for the photometric double lensing events and is represented by
\begin{equation}
-\alphas \le \alpha \le \alphao;
\end{equation}
\begin{displaymath}
\alphao = \sin^{-1} \left( \uthf + \sqrt{q}\uths \over d\right ),
\end{displaymath}
\begin{displaymath}
\ \ \alphas = \sin^{-1} \left( |\uthf - \sqrt{q}\uths| \over d\right ),
\end{displaymath}
for $d > \uthf + \sqrt{q}\uths$,
\begin{equation}
-\alphas \le \alpha \le {\pi \over 2},
\end{equation}
for $|\uthf - \sqrt{q}\uths| < d \le \uthf + \sqrt{q}\uths$, and
\begin{equation}
-{\pi \over 2} \le \alpha \le {\pi \over 2},
\end{equation}
for $d \le |\uthf - \sqrt{q}\uths|$.
Here $\alpha$ is the angle between the source trajectory and the binary axis, $d$ is the projected binary separation normalized to the Einstein radius of the primary, $\uthf$ and $\uths$ are the threshold impact parameters to the primary and the companion normalized to the Einstein ring radii of the individual lens components \citep{han02}.
The indices $``\rm o"$ and $``\rm s"$ represent the cases when the impact parameters are located on the opposite and same sides to the binary axis, respectively.

\section{NEW METHOD FOR THE DEGENERACY}

For resolving the close/wide degeneracies in binary lensing events using a new astrometric method, we use two solutions of the MACHO 98-SMC-1 event that are very similar in terms of the light curve and centroid shift (Afonso et al. 2000), but which have opposite source trajectories.
Figure 1 shows the geometries of the close and wide solutions of a severely degenerate binary lensing event.
In Figure 1, $s$ is the projected binary separation and $q$ is the mass ratio of the binary.
The coordinates $(\xi,\eta)$ represent the axes that are parallel with and normal to the binary axis.
The coordinates for the close and wide binaries are centered at the center of mass and the midpoint of the binary lens, respectively.
Here all lengths in the close and wide binaries are normalized to the individual Einstein ring radii of the two cases.
The straight line with an arrow represents the source trajectory and the curve that varies with the source trajectory represents the trajectory of the image centroid. The solid curve represents the image centroid dominated by the companion, $m_2$, while the dot curve represents the image centroid dominated by the primary, $m_1$.
The dashed circles for the wide binary represent the Einstein radii of the individual lens components, while for the close binary the dashed circle represents the combined Einstein radius of the binary.
Two dots are the locations of the individual lens components.
The concave curves with cusps located around or away from the lenses are caustics that represent the set of source positions at which the Jacobian determinant of the lens equation becomes zero, and thus the magnification of a point source becomes infinite.
A plus sign in the caustic represents the center of mass of the binary for the close binary and represents the effective position for the wide binary.

The criterion for which lens of the binary lens components has a greater effect on the binary lensing event is determined by
\begin{equation}
u_{i,\rm eff} = \sqrt{(x_s - x_{i,\rm eff})^2 + (y_s - y_{i,\rm eff})^2}\ ,
\end{equation}
where $u_{i,\rm eff}$, $(x_s ,y_s)$, and $(x_{i,\rm eff},y_{i,\rm eff})$ represent the effective lens-source separation, the source position, and the effective positions of the individual lens components, respectively, which are normalized to the combined Einstein radius of the binary lens system.
Therefore, the solid curve in Figure 1 is for $u_{1,\rm eff} > u_{2,\rm eff}$, while the dotted curve is for $u_{1,\rm eff} < u_{2,\rm eff}$.
The light curves and the centroid shift trajectories resulting from the source trajectories represented in Figure 1 are shown in Figure 2.
In Figure 2, $\theta_{\rm E,c}$ and $\theta_{\rm E,w}$ represent the Einstein radii corresponding to the total masses of the close and wide binaries, respectively.
An arrow in the lower panel represents the orientation of the centroid shift trajectory.
The dot curve in the left side of the lower panel is caused by the primary.
Since the first observed event is caused by the companion in the wide binary, the additional deviation in the centroid shift trajectory is caused by the primary.
The additional deviation can be induced by the primary or the companion, depending on which lens of the binary lens components causes the first observed event.
Assuming that the first event is a typical Galactic lensing event caused by $m = 0.3 \ M_\odot$ and with $D_{\rm S} = 8 \ \rm kpc$ and $D_{\rm L} = 6 \ \rm kpc$, the Einstein timescale is $\te \sim 20$ days, the angular Einstein radius is $\thetae \sim 300\ \mu \rm as$, and the centroid shift is usually several tens of $\mu \rm as$.
Under this assumption, the crosses marked in Figure 2 show the progress of the event in the interval of 30 days after the caustic-crossing.
As shown in the figure, the additional deviation appears about 30 days after the caustic-crossing and the centroid shift at the position of the next cross changes by $\sim\ 152\ \mu \rm as$ from the first cross position.
If this remarkable additional deviation is detected by using future astrometric follow-up observations, such as \emph{SIM PlanetQuest} with the position accuracy of $4\ \mu \rm as$, the degeneracy in binary microlensing events can be easily resolved.
However, if the source passes too far away from the Einstein ring of the lens causing the second event, it will be difficult to detect the additional deviation.

The source should cross right on the caustic or pass in the vicinity of the caustic to be securely identified as binary lensing events, so that most of the observed binary lensing events are high magnification events.
To determine the probability of detecting the remarkable additional centroid shift of the second event in wide binary lensing events with high magnification,
we set $\uthf = 0.1$ and $\uths = 2.5$ in equations $(6)-(8)$.
Because of the long range astrometric effect of the lens causing the second event, the additional deviation presents well in the centroid shift trajectory, despite the fact that $\uths$ is 2.5 times larger than the Einstein ring of the lens.
We also set that the lens causing the first event is the primary and the lens causing the second event is the companion. In this case, since the mass of the companion causing the second event can be heavier than that of the primary, the binary mass ratio can be larger than $q = 1.0$.
The probability is determined by
\begin{equation}
P = {\alphao + \alphas \over {\pi}},
\end{equation}
and the result is presented in Figure 3.
As shown in Figure 3, the probability increases as the binary mass ratio increases and the binary separation decreases, because the efficient astrometric range increases as the mass ratio increases and the separation decreases.
The separation distribution of three of the four binary lensing events with known close and wide solutions is $d \lesssim 10.0$.
We thus present the probability under the two different binary separation distributions in Table 1.
The physical separation is determined by assuming that the physical Einstein radius of the typical Galactic lensing event is $r_{\rm E,1} \sim 2 \rm \ AU$.
From the determination, we find that in the case where the mass of the companion is 1.5 times heavier than that of the primary, i.e., $q = 1.5$, the degeneracy of binary lensing events with a separation of $\lesssim 20 \rm \ AU$ can be resolved with an efficiency $\sim 30 \%$.

The issue of how long we should wait for the detection of the centroid shift of the second event after the caustic-crossing of the first event in wide binary lensing events depends on both the binary separation and the Einstein timescale of the first event for a given binary mass ratio.
The $``\rm waiting \ time"$ for the detection of the additional centroid shift in wide binary lensing events decreases as the binary separation and the Einstein timescale of the first event decrease.
When the caustic-crossing events are detected, various lensing parameters can be estimated from the fit of those light curves.
The separation and the Einstein timescale of them would be helpful to foresee the waiting time.
For predicting the waiting time, we estimate it by using the distributions of the separation and the Einstein timescale of observed Galactic lensing events.
Figure 4 represents the waiting time for wide binary lensing events with $q = 1.5$.
From this, we find that for typical Galactic events with $t_{\rm E,1} \sim 20$ days at a separation of $\lesssim 20.0\ \rm AU$, one can detect the additional centroid shift within $100$ days.

\section{SUMMARY AND DISCUSSIONS}

We have investigated whether the close/wide degeneracies in binary microlensing events can be resolved by detecting the additional centroid shift induced by the lens causing the second event in wide binary microlensing events.
From this investigation, we found that in the case where the source passes close to the Einstein ring of the lens causing the second event, the degeneracy could be easily resolved by using future astrometric follow-up observations.
We have determined the probability of detecting the additional centroid shift in high magnification binary lensing events.
From this, we found that the degeneracy of binary lensing events with the separation of $\lesssim 20\ \rm AU$ could be resolved with a significant efficiency.
We have also estimated the waiting time for the detection of the additional centroid shift in wide binary lensing events.
From this, we found that for typical Galactic binary lensing events, the degeneracy could be sufficiently resolved within a year.

The targets of \emph{SIM PlanetQuest} are usually limited to rather bright source stars and thus the degeneracy for very faint sources with $V > 20$ could not be broken by this mission. The probability presented in Sec. 3 has not considered the source brightness; if the source brightness were to be considered, then the probability would actually more decrease.
Considering that for the Galactic lensing events with $t_{\rm E,1} \sim 20$ days the additional deviation would appear about a month after the caustic-crossing and the degeneracy could be broken within 100 days, the astrometric observations once a month would be enough for breaking the close/wide degeneracy.
Unfortunately, the mission of \emph{SIM PlanetQuest} has been deferred and there are also no other astrometric missions with the capabilities of \emph{SIM}, while in the near future the ground-based photometric microlensing survey would be carried out for 24-hour by the global telescope network with three 1.6m class telescopes located at good seeing sites (\emph{Korea Microlensing Telescope Network} (\emph{KMTNet}) Project \citep{han09}).
Recently, the newly redesigned \emph{SIM-Lite}, which costs less than the original mission, and which has improved, has been proposed by \citet{shao08}.
However, even though the mission is due to restart in the near future, it is very difficult to foresee its launching date.

\begin{figure}[t]
\epsscale{1.0}
\plotone{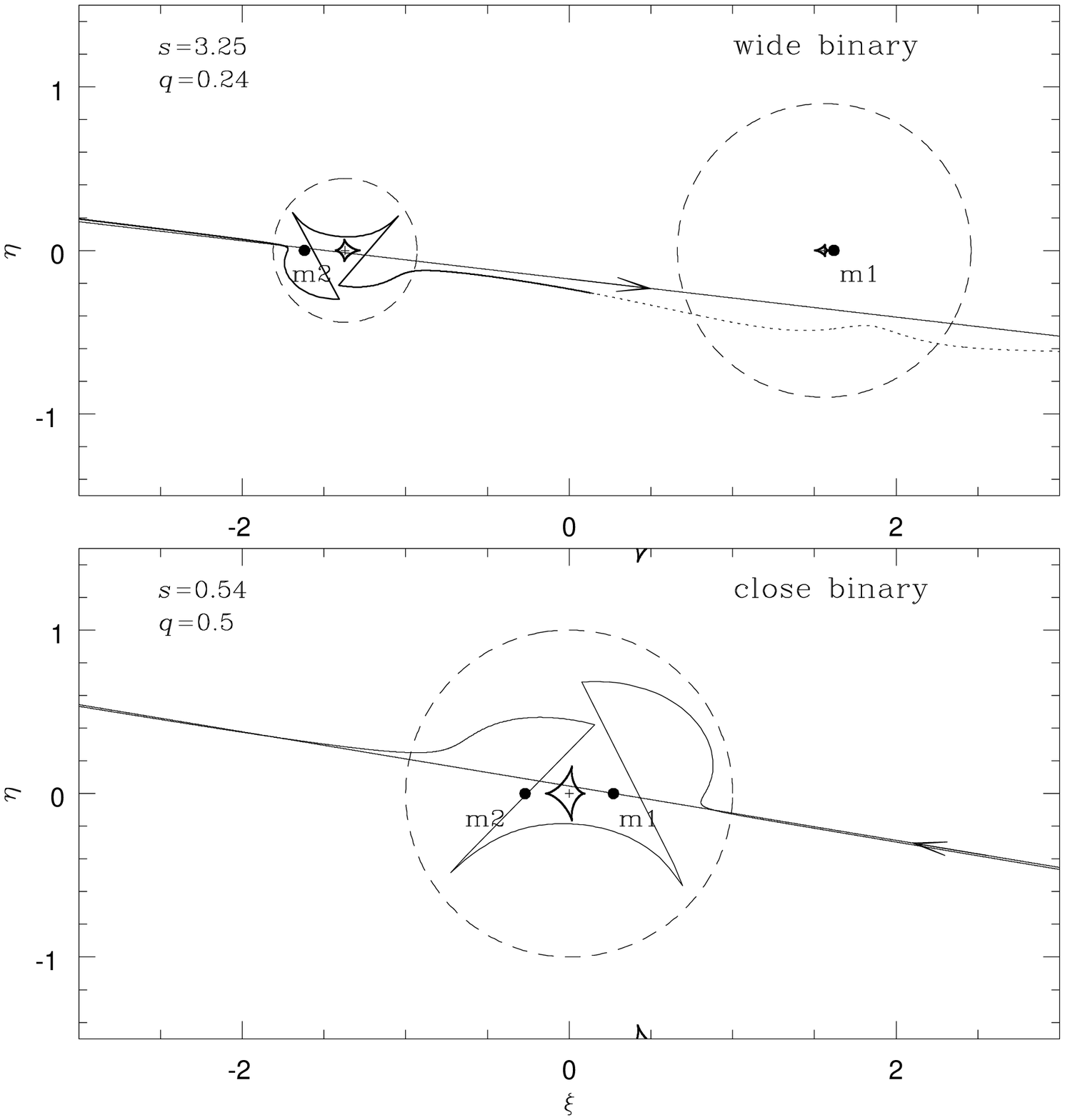}
\caption{\label{fig:one}
The geometries of the close and wide solutions of a severely degenerate binary lensing event.
Lensing parameters $s$ and $q$ are the projected separation and the mass ratio of the binary, respectively.
The straight line with an arrow represents the source trajectory and the curve that varies with the source trajectory represents the trajectory of the image centroid.
The solid and dotted curves represent the image centroid induced by the companion, $m_2$, and primary, $m_1$, respectively.
The coordinates $(\xi,\eta)$ represent the axes that are parallel with and normal to the binary axis. The coordinates for the close and wide binaries are centered at the center of mass and the midpoint of the binary lens, respectively.
Here all lengths in the close and wide binaries are normalized to the individual Einstein radii of two cases.
}\end{figure}

\begin{figure}[t]
\epsscale{1.0}
\plotone{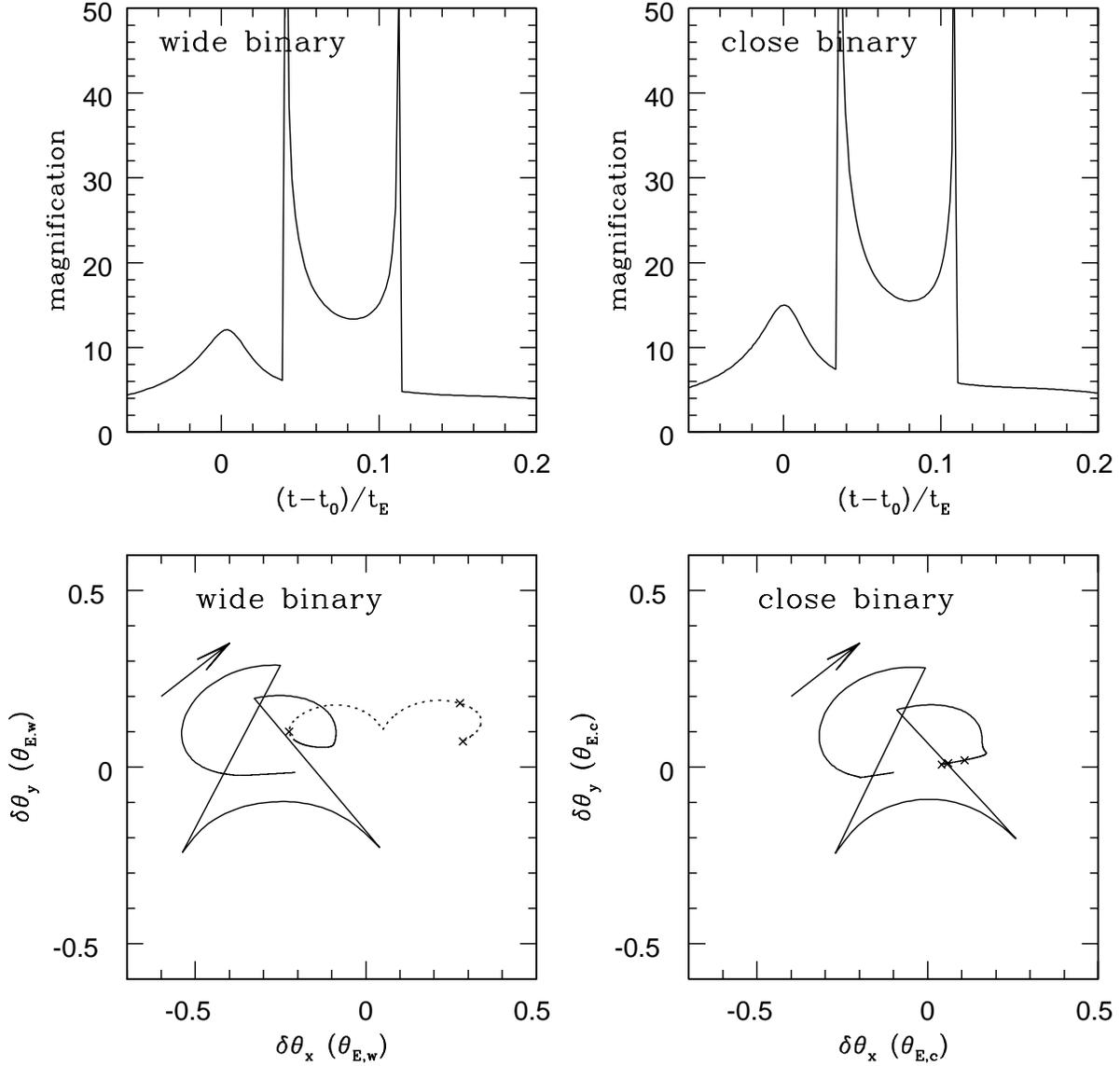}
\caption{\label{fig:two}
The light curve and centroid shift trajectory of the source images for the
binary lensing events represented in the Fig. 1. The arrow in the lower panel represents the direction of the centroid shift trajectory. The dotted curve in the left side of the lower panel is the centroid shift caused by the primary as shown in Fig. 1. In the figure, $\theta_{\rm E,c}$ and $\theta_{\rm E,w}$ represent the Einstein ring radii corresponding to the total masses of the close and wide binaries, respectively. The crosses show the progress of the event in the interval of 30 days after the caustic-crossing.
}\end{figure}

\begin{figure}[t]
\epsscale{1.0}
\plotone{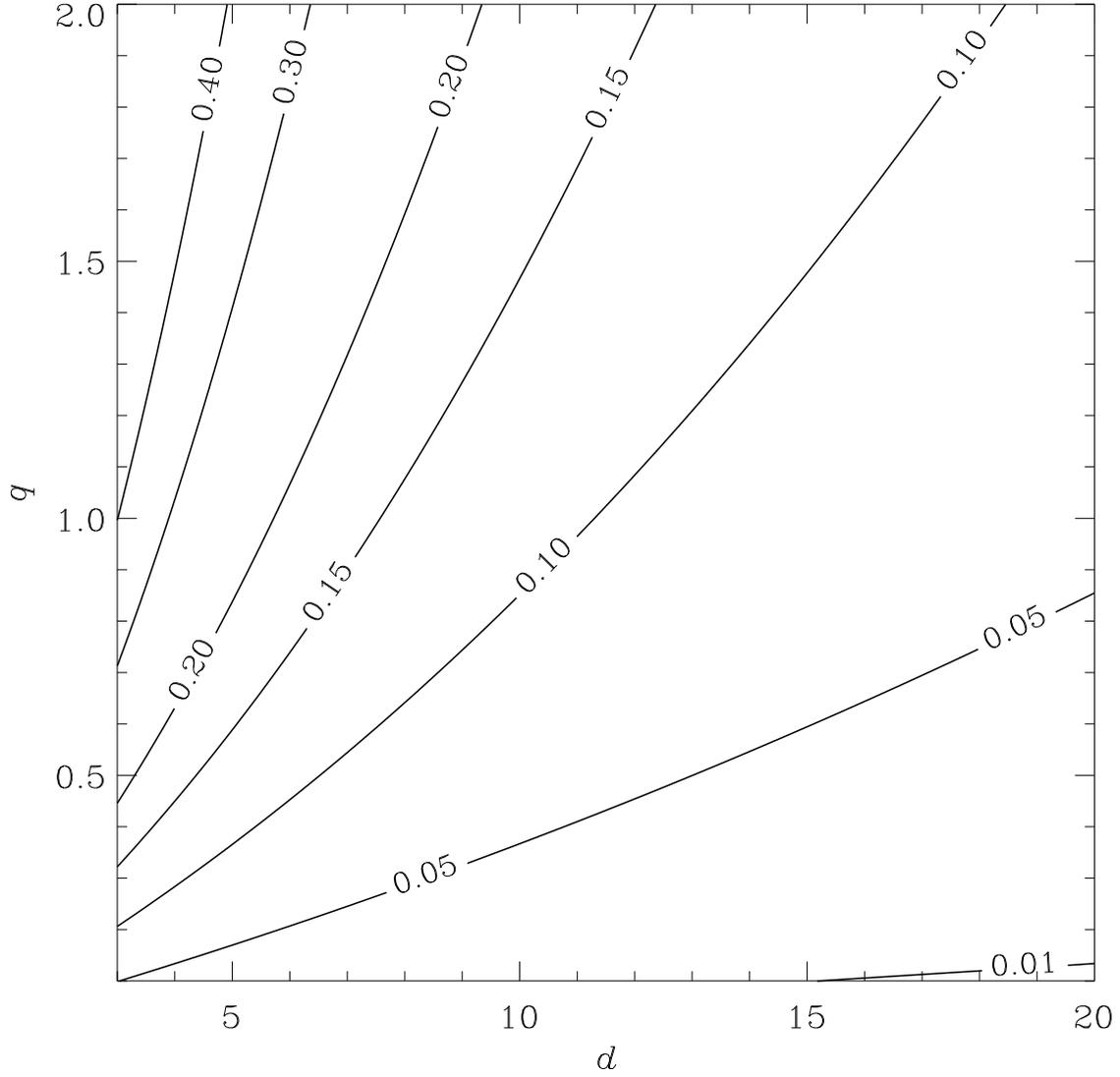}
\caption{\label{fig:three}
Probability of detecting the additional centroid shift induced by the lens causing the second event in wide binary lensing events as a function of the separation and mass ratio of the binary. In the figure, $d$ represents the projected binary separation normalized by the Einstein ring radius of the primary causing the first event.
}
\end{figure}

\begin{figure}[t]
\epsscale{1.0}
\plotone{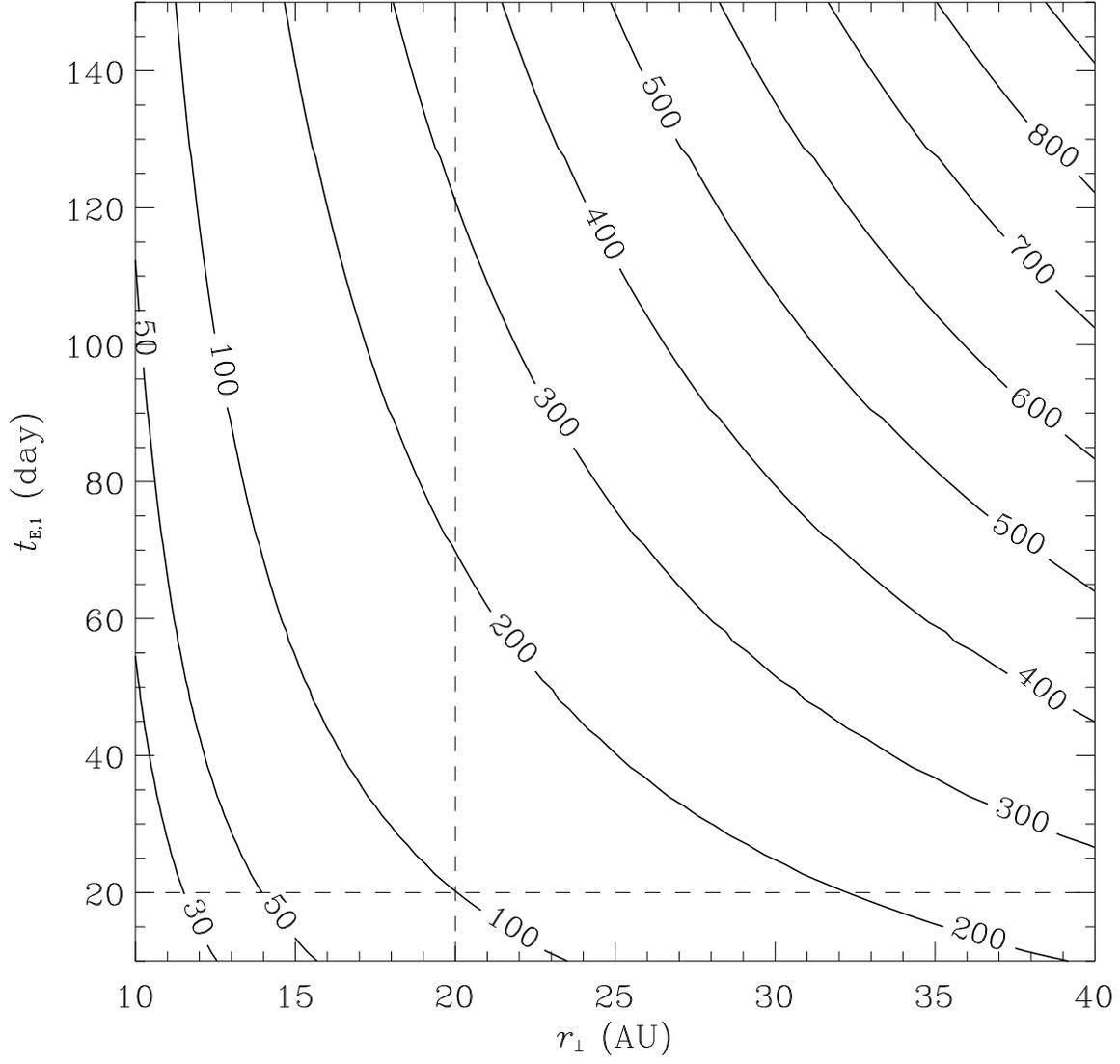}
\caption{\label{fig:four}
The waiting time for the detection of the centroid shift of the second event after the caustic-crossing of the first event in wide binary lensing events with $q = 1.5$ as a function of the binary separation and Einstein timescale of the first event.
The unit of numbers in the contour is day.
}
\end{figure}

\begin{deluxetable}{ccc}
\tablecaption{Probability of detecting the additional centroid shift.\label{tbl-one}}
\tablewidth{0pt}
\tablehead{
&\multicolumn{2}{c} {Probability ($\%$)} \\ \cline{2-3} \\
Binary mass ratio & $6\ {\rm AU} \lesssim r_{\perp} \lesssim 40\ {\rm
AU}$ & $6\ {\rm AU} \lesssim r_{\perp} \lesssim 20\ {\rm AU}$ }
\startdata
0.5 & 7.7 & 11.3 \\
1.0 & 13.6 & 19.8 \\
1.5 & 18.5 & 27.2 \\
\enddata
\tablecomments{
Probability of detecting the remarkable centroid shift of the second event in wide binary lensing events with the two different distributions of the binary separation. The physical separation, $r_\perp$, is obtained by assuming a typical lensing event with $D_{\rm S} = 8 \ {\rm kpc}$, $D_{\rm L} = 6 \ {\rm kpc}$, and $m_{1} = 0.3\ M_\odot$.
}
\end{deluxetable}


\end{document}